\documentclass[A4,10pt,twocolumn]{article} 
\usepackage[american]{babel}
\usepackage{amsmath} 
\usepackage{pxfonts}
\usepackage{graphicx}
\usepackage{times}
\usepackage{fullpage}
\usepackage{color}

\definecolor{lightgray}{gray}{0.95}
\definecolor{darkgray}{gray}{0.65}

\newcommand{\equals}{\stackrel{\mathrm{def}}{=}}

\newcommand{\mode}{M}
\newcommand{\sched}{S}
\newcommand{\MCR}{\operatorname{MCR}}
\newcommand{\upmakespan}{\operatorname{upms}}
\newtheorem{Corollary}{Corollary}
 
\newtheorem{Lemma}{Lemma}

\newtheorem{Definition}{Definition}

\newtheorem{Condition}{Schedulability Condition}

\newenvironment{proof}[1][Proof]{\begin{trivlist}
\item[\hskip \labelsep {\bfseries #1}]}{\end{trivlist}}

\newcommand{\qed}{\nobreak \ifvmode \relax \else
      \ifdim\lastskip<1.5em \hskip-\lastskip
      \hskip1.5em plus0em minus0.5em \fi \nobreak
      \vrule height0.75em width0.5em depth0.25em\fi}

\title{Mode Change Protocol for Multi-Mode Real-Time Systems\\
 upon Identical Multiprocessors}
\author{Vincent N\'elis$^{1,2}$, Jo\"el Goossens$^1$ \\
\{vnelis, joel.goossens\}@ulb.ac.be\\
}

\begin{document}
\date{}
\maketitle
\footnotetext[1]{Universit\'e Libre de Bruxelles (U.L.B.), Facult\'e des Sciences, Scheduling Group, Brussels, Belgium.}
\addtocounter{footnote}{1}
\footnotetext[2]{Supported by the Belgian National Science Foundation (FNRS) under a FRIA grant.}
\addtocounter{footnote}{1}

\begin{abstract}
In this paper, we propose a synchronous protocol without periodicity for scheduling multi-mode real-time systems upon identical multiprocessor platforms. Our proposal can be considered to be a multiprocessor extension of the uniprocessor protocol called ``Minimal Single Offset protocol''.
\end{abstract}

\section{Introduction}

Hard real-time systems require both functionality correct executions and results that are produced on time. Control of the traffic (ground or air), control of engines, control of chemical and nuclear power plants are just some examples of such systems. Currently, numerous techniques exist that enable engineers to design real-time systems, while guaranteeing the correctness of their temporal behavior in a systematic way. These techniques generally model each functionality of the system by a \emph{recurrent} task, characterized by a computing requirement, a temporal deadline and an activation rate. Commonly, real-time systems are modeled by a fixed set of such tasks. However, some applications exhibit multiple behaviors issued from several operating modes (e.g.,~an initialization mode, an emergency mode, a fault recovery mode, etc.), where each mode is characterized by its own set of functionalities, i.e., its set of tasks. During the execution of such \emph{multi-mode} real-time systems (described in details in~\cite{JoAlfons:04}), switching from the current mode (called the \emph{old-mode}) to another one (the \emph{new-mode} hereafter) requires to substitute the current executing tasks with the tasks of the target mode. This substitution introduces a transient stage, where the tasks of the old- and new-mode may be scheduled \emph{simultaneously}, thereby leading to an overload which compromises the system schedulability.

The scheduling problem during a transition between two modes has multiple aspects, depending on the behavior and requirements of the old- and new-mode tasks when a mode change is initiated (see e.g., ~\cite{Fohler:94,Jahanian:88}). For instance, an old-mode task may be immediately aborted, or it may require to complete the execution of its current instance in order to preserve data consistency. On the other hand, a new-mode task sometimes requires to be activated as soon as possible, or it may also have to delay its first activation until all the tasks of the old-mode are completed. Moreover, it may exist some tasks (called \emph{mode-independent} tasks) present in both the old- and new-mode, such that their periodic executions must be carried out independently from the mode change in progress. The existing transition scheduling protocols are classified with respect to their ability to deal with the mode-independent tasks and the way old- and new-mode tasks are handled during the transitions. In the literature (see~\cite{JoAlfons:04} for instance), a protocol is said to be \emph{synchronous} if it never releases the new-mode tasks before \emph{all} the old-mode tasks have completed their last instance, otherwise it is said to be \emph{asynchronous}. Furthermore, a synchronous/asynchronous protocol is said to be \emph{with periodicity} if it is able to deal with mode-independent tasks, otherwise it is said to be \emph{without periodicity}.

\noindent \textbf{Related work.} Numerous scheduling protocols have already been proposed in the \emph{uni}processor case to ensure the transition between two modes. In synchronous protocols, one can cite the \emph{Idle Time Protocol}~\cite{Tindell:96} where the periodic activations of the old-mode tasks are suspended at the first idle time-instant occurring during the transition and then, the new-mode tasks are released. The~\emph{Maximum-Period Offset Protocol} proposed in~\cite{Bailey:93} is a protocol \emph{with periodicity} which delays the first activation of all the new-mode tasks for a time equal to the period of the less frequent task in both modes (mode-independent tasks are not affected). The \emph{Minimum Single Offset Protocol} in~\cite{JoAlfons:04} completes the last activation of all the old-mode tasks and then, releases the new-mode ones. This protocol exists in two versions, \emph{with} and \emph{without periodicity}. Concerning the \emph{asynchronous} protocols, the authors of~\cite{Tindell:92} propose a protocol \emph{with periodicity} and the authors of~\cite{Pedro:98,Pedro:99} propose a protocol \emph{without periodicity}. 

To the best of our knowledge, no protocol exist in the \emph{multi}processors case. This problem is much more complex, especially due to the presence of scheduling anomalies upon multiprocessors (see, e.g., chapter 5 of~\cite{Andersson:03} on page 51 for a definition). Nowadays, it is well-known that real-time multiprocessor scheduling problems are typically not solved by applying straightforward extensions of techniques used for solving similar uniprocessor problems. \\

\noindent \textbf{This research.} In this paper, we propose a \emph{synchronous protocol without periodicity} for the identical multiprocessor case. Our proposal can be considered to be a multiprocessor extension of the Minimal Single Offset protocol, proposed in~\cite{JoAlfons:04}. \\

\noindent \textbf{Paper organization.} In Section~\ref{sec_model}, we define the computational model used throughout the paper. In Section~\ref{sec_protocol}, we propose a \emph{synchronous} protocol \emph{without periodicity} for the identical multiprocessor case, and in Section~\ref{sec_futurework}, we introduce future research directions.

\section{Model of computation}
\label{sec_model}
We consider a multiprocessor platform composed of $m$ identical processors denoted by ${\cal P}_1$, ${\cal P}_2$, $\ldots$, ${\cal P}_m$. We define a multi-mode real-time system $\tau$ as a set of $x$ operating modes noted $\mode_1, \mode_2, \ldots, \mode_x$ where each mode contains its own set of functionalities to execute. At any time during its execution the system runs in only one of its modes, meaning that it executes only the set of tasks associated with the selected mode.

A mode $\mode_k$ contains a set $\tau^k$ of $n_k$ functionalities, each modeled by a sporadic task $\tau_i^k = (C_i^k, D_i^k, T_i^k)$, defined by three parameters -- a worst-case execution time $C_{i}^k$, a minimal inter-arrival delay $T_{i}^k$ and a deadline $D_{i}^k \leq T_i^k$ -- with the interpretation that the task generates successive \emph{jobs} $\tau_{i,j}^k$ (with $j = 1, \ldots, \infty$) arriving at times $e_{i,j}^k$ such that $e_{i,j}^k \geq e_{i,j-1}^k + T_i^k$ (with $e_{i,1}^k \geq 0$), each such job has an execution requirement of at most $C_{i}^k$, and must be completed at (or before) its deadline noted $D_{i,j}^k \equals e_{i,j}^k + D_i^k$. Since $D_i^k \leq T_i^k$, successive jobs of a task $\tau_i^k$ do not interfere with each other. Notice that a task must be \emph{enabled} to generate jobs, and the system is said to run in mode $\mode_k$ only if every task of $\tau^k$ is enabled and all the tasks of the other modes are disabled. Thereby, disabling a task prevents it from releasing its future jobs. In our study, all the tasks of every mode are assumed to be independent, i.e., there is no communication, no precedence constraint and no shared resource (except the processors) between them. 

Every mode $\mode_k$ of the system uses its own scheduling algorithm noted $\sched_k$, which must guarantee that all the deadlines of the tasks of $\tau^k$ are met, when executed upon the $m$ processors. In our study, we assume that $\sched_k$ is (1) \emph{global}, i.e., a job may be assigned to any processor, (2) \emph{conservative}, i.e., a processor cannot be idle if there is a pending job, and (3) must assign \emph{fixed job-level priorities}, i.e., a job gets a priority as soon as it is released and keeps it constant until it completes. At run-time the scheduler assigns, at each time-instant, the $m$ highest priority jobs (if any) to the $m$ processors. Global-EDF and Global-DM (see e.g., ~\cite{Baker:03}) are some examples of such schedulers.

While the system is running in a mode $\mode_i$, a mode change can be initiated by any task of $\tau^i$ or by the system itself, whenever it detects a change in the environment or in its internal state. This is performed by releasing a $\MCR(j)$ (i.e., a Mode Change Request), where $\mode_j$ is the targeted destination mode. In the following, we denote by $t_{\MCR(j)}$ the releasing time of a $\MCR(j)$. We assume that a MCR may only be produced in the steady state of the system, and not during the transition between two modes. Upon a $\MCR(j)$, the old-mode tasks (i.e., the tasks of the current mode $\mode_i$) may have two distinct behaviors: either they must be aborted, or they have to complete the execution of their last released job. We denote by $C(i,j)$ the subset of tasks of $\tau^i$ which must complete their last instance when a $\MCR(j)$ occurs (and therefore $\tau^i \setminus C(i,j)$ is the subset of tasks of $\tau^i$ which can be aborted). Notice that we do not consider mode-independent tasks in this paper. 

Whenever a $\MCR(j)$ occurs, the system immediately entrusts the scheduling decisions to a transition protocol. This protocol aborts all the task of $\tau^i \setminus C(i,j)$ and it disables all the tasks (if any) of $C(i,j)$. Nevertheless, the last released job of every task of $C(i,j)$ is not aborted and must complete its execution. We call these jobs the \emph{rem-jobs} hereafter. Since these rem-jobs may cause an overload if the tasks of $\tau^j$ are immediately enabled upon a $\MCR(j)$, the transition protocol usually has to delay the enablement of these tasks until it is safe to enable them. We denote by ${\cal D}(k, \mode_i, \mode_j)$ the relative deadline on the enabling time of the task $\tau_k^j$ during the transition from the mode $\mode_i$ to the mode $\mode_j$, with the following interpretation: the transition protocol must ensure that $\tau_k^j$  is not enabled after time $t_{\MCR(j)} + {\cal D}(k, \mode_i, \mode_j)$. The goal of a transition protocol is therefore to complete every rem-job and to enable every task of the mode $\mode_j$, while meeting every job and enablement deadline. When all the rem-jobs are completed and all the tasks of $\tau^j$ are enabled, the system entrusts the scheduling decisions to the scheduler $\sched_j$ of the mode $\mode_j$ and the transition phase ends.

\section{Multiprocessor synchronous protocol without periodicity}
\label{sec_protocol}
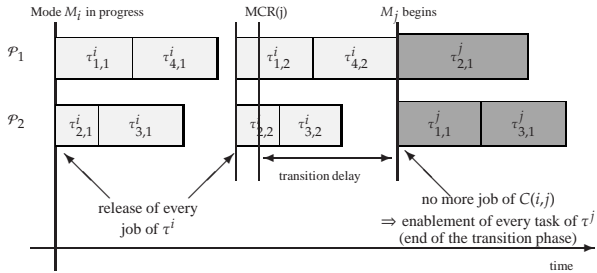
\begin{figure}[h!]
\centering
  {\tiny \setlength{\unitlength}{0.3cm}
\begin{picture}(25,12)

\put(1,0){\vector(1,0){25}}
\put(24,-1){time}

\put(0,8.5){${\cal P}_1$}
\put(0,5.5){${\cal P}_2$}

\put(2.1,-1){\line(0,1){11}}
\put(1,10.3){Mode $\mode_i$ in progress}

\put(2.1,8.3){\fcolorbox{black}{lightgray}{\makebox[3\unitlength][c]{$\tau_{1,1}^{i}$}}}
\put(5.5,8.3){\fcolorbox{black}{lightgray}{\makebox[3\unitlength][c]{$\tau_{4,1}^{i}$}}}
\put(2.1,5.3){\fcolorbox{black}{lightgray}{\makebox[1.5\unitlength][c]{$\tau_{2,1}^{i}$}}}
\put(4,5.3){\fcolorbox{black}{lightgray}{\makebox[3\unitlength][c]{$\tau_{3,1}^{i}$}}}

\put(10.1,8.3){\fcolorbox{black}{lightgray}{\makebox[3\unitlength][c]{$\tau_{1,2}^{i}$}}}
\put(13.5,8.3){\fcolorbox{black}{lightgray}{\makebox[3.1\unitlength][c]{$\tau_{4,2}^{i}$}}}
\put(10.1,5.3){\fcolorbox{black}{lightgray}{\makebox[1.5\unitlength][c]{$\tau_{2,2}^{i}$}}}
\put(12,5.3){\fcolorbox{black}{lightgray}{\makebox[2\unitlength][c]{$\tau_{3,2}^{i}$}}}
\put(11.1,3){\line(0,1){7}}
\put(10.5,10.3){$\operatorname{MCR(j)}$}

\put(17.25,8.3){\fcolorbox{black}{darkgray}{\makebox[5\unitlength][c]{$\tau_{2,1}^{j}$}}}
\put(17.25,5.3){\fcolorbox{black}{darkgray}{\makebox[3\unitlength][c]{$\tau_{1,1}^{j}$}}}
\put(20.95,5.3){\fcolorbox{black}{darkgray}{\makebox[3\unitlength][c]{$\tau_{3,1}^{j}$}}}
\put(17.25,3){\line(0,1){7}}
\put(16.5,10.3){$\mode_j$ begins}

\put(4.5, 2.2){\vector(-1,1){2}}
\put(8, 2.2){\vector(1,1){2}}
\put(10.1,3){\line(0,1){7}}
\put(4,1){{\footnotesize $\substack{\text{release of every}\\ \text{job of $\tau^i$}}$}}

\put(14,4){\vector(1,0){3}}
\put(14,4){\vector(-1,0){2.7}}
\put(12,3){{\tiny transition delay}}

\put(19.6, 2.3){\vector(-1,1){2}}
\put(16.5,1){{\footnotesize $\substack{\text{no more job of } C(i,j) \\ \Rightarrow\ \text{enablement of every task of } \tau^j \\ \text{(end of the transition phase)} }$}}

\end{picture}
}
\caption{Illustration of a mode transition handled by our synchronous protocol.}
\label{F_example}
\end{figure}

In this section, we present our synchronous protocol without periodicity. The main idea is the following: \emph{upon a $\MCR(j)$, the rem-jobs continue to be scheduled by $\sched_i$ upon the $m$ processors. When all of them are completed, all the new-mode tasks are simultaneously enabled}. Figure~\ref{F_example} depicts an example with a 2-processors platform. The modes $\mode_i$ and $\mode_j$ contain 4 and 3 tasks, respectively. The light and dark gray boxes are the old- and new-mode tasks, respectively. In this example, we consider that every task of $\tau^i$ belongs to $C(i,j)$. 

\begin{Definition}[a valid protocol]
A transition scheduling protocol is said to be \emph{valid} for a given multi-mode real-time system if it always meets all the job and enablement deadlines during the transition from any mode of the system to any other one.
\end{Definition}

In order to know if a our protocol is valid for a given multi-mode system, we need to establish a \emph{schedulability condition}, i.e., a condition based on the tasks and platform characteristics which indicates \emph{a priori} whether the given system will always comply with the expected requirements during its execution. In the following, we first proof that every rem-job \emph{always} meets its deadline during a transition from a mode $\mode_i$ to another mode $\mode_j$, when scheduled by $\sched_i$ upon the $m$ processors. Then, we express a \emph{sufficient} schedulability condition which indicates if all the enablement deadlines are met during every mode change, while considering the worst-case scenario. 

\begin{Lemma} 
\label{lemma2}
When a $\MCR(j)$ occurs at time $t_{\MCR(j)}$ while the system is running in mode $\mode_i$, every rem-job issued from the tasks of $C(i,j)$ meets its deadline when scheduled by $\sched_i$ upon $m$ processors. 

\begin{proof}
From our assumptions, we know that the set of tasks $\tau^i$ is schedulable by $\sched_i$ upon $m$ processors, and since $\sched_i$ assigns fixed job-level priorities, we know from~\cite{Ha1994} that $\sched_i$ is \emph{predictable}. By definition of the predictability (see~\cite{Ha1994} for details), if a set of jobs $J = \{ J_1, J_2, \ldots, J_n \}$ (where each job $J_i = (a_i, c_i, d_i)$ is characterized by an arrival time $a_i$, a computing requirement $c_i$ and a absolute deadline $d_i$) meets all the deadlines when scheduled by a predictable algorithm $A$ upon $m$ processors, then any set of jobs $J' = \{ J_1', J_2', \ldots, J_n' \}$, where $J_i' = (a_i, c_i', d_i)$ with $c_i' \leq c_i$, also meets all the deadlines when scheduled by $A$ upon the $m$ processors. When the $\MCR(j)$ occurs at time $t_{\MCR(j)}$, every task of $\tau^i \setminus C(i,j)$ is aborted and every task of $C(i,j)$ is disabled. Aborting and disabling these tasks is equivalent to set the execution requirement of all their jobs (or future jobs for the disabled tasks) to zero. Since $\sched_i$ is predictable, all the deadlines are still met when the rem-jobs are scheduled by $\sched_i$ upon the $m$ processors.
\qed
\end{proof}
\end{Lemma}

From Lemma~\ref{lemma2}, every rem-job always meets its deadline when using our proposed protocol during the transition. Thereby, our protocol is valid for a given real-time system if, for every mode transition, the maximal delay which may be produced by the rem-jobs is lower or equal than the enablement deadline of every new-mode task. That transition delay is equal to the completion time (also called \emph{makespan} in the literature and hereafter) of all the rem-jobs and hence, we establish in the following lemma an upper bound on the makespan of a given set of $n$ jobs $J = \{ J_1, J_2, \ldots, J_n\}$ of processing time $p_1, p_2, \ldots, p_n$. 

\begin{Lemma}
\label{lemma1}
Let $J = \left\{ J_1, J_2, \ldots, J_n \right\}$ be a set of $n$ jobs with processing times $p_1, p_2, \ldots, p_n$ which are ready for execution at time $0$. Suppose that these jobs are scheduled upon $m$ identical processors (with $m > 1$) by a global, conservative and fixed job-level priority scheduler. Whatever the jobs priority assignment, an upper bound on the makespan is given by:
\begin{small}
\begin{equation}
\label{upmakespan}
\upmakespan(J,m) = 
\begin{cases}
p_{\max} & \text{if } m \geq n \\
\frac{1}{m} \sum_{i=1}^n p_i + (1 - \frac{1}{m})\ p_{\max} & otherwise
\end{cases}
\end{equation}
\end{small}
\noindent where $p_{\max} \equals \max_{i=1}^n\{ p_i \}$. 
\end{Lemma}

The proof is omitted in this version of the paper due to the space limitation.

\begin{Corollary}
\label{corollary1}
For any job $J_i$ of processing time $p_i$ and for any set of jobs $J$, we have $\upmakespan(J \cup \{ J_i \}, m) \geq \upmakespan(J, m)$.
\end{Corollary}

\begin{Corollary}
\label{corollary2}
For any set of jobs $J$ and for any job $J_i \in J$, let $J_i'$ be a job such that $p_i' \geq p_i$. We have $\upmakespan(J \setminus \{J_i\} \cup \{ J_i' \}, m) \geq \upmakespan(J, m)$.
\end{Corollary}

In the framework of our problem, we know from Corollary~\ref{corollary1} and~\ref{corollary2} that the worst-case scenario occurs when (i) every task of $C(i,j)$ releases a job exactly at time $t_{\MCR(j)}$ and (ii) every released job has a processing time equal to the worst-case execution time of its task. Thus, a \emph{sufficient} schedulability condition may be formalized as follows: 

\begin{Condition}
\label{condition1}
For every transition from a mode $M_i$ to another mode $M_j$, it must be the case where:
 \[ \upmakespan(J, m) \leq \min_{k=1}^{n_j} \{ {\cal D}(k, \mode_i, \mode_j) \} \]
\noindent where $J \equals \{ C_r^i \mid \tau_r^i \in C(i,j) \}$ and $\upmakespan(J, m)$ is defined by Equation~\ref{upmakespan}.
\end{Condition}

\noindent \textbf{Open Problem 1 } Instead of scheduling the rem-jobs by using the scheduler of the old-mode during the transition, it could be better, in term of the enablement delays applied to the new-mode tasks, to propose a \emph{dedicated priority assignment} which meets the deadline of every rem-job, \emph{while} minimizing the makespan. To the best of our knowledge, the problem of minimizing the makespan while meeting job deadlines together is not addressed in the literature and remains for future work.

\section{Future work}
\label{sec_futurework}

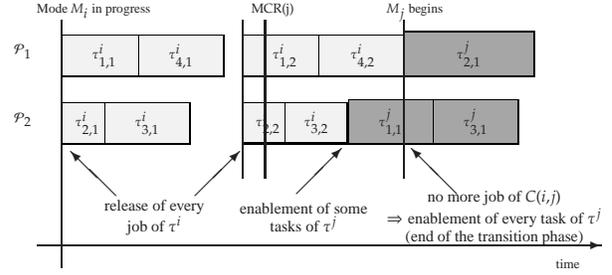
\begin{figure}[h!]
\centering
  {\tiny \setlength{\unitlength}{0.3cm}
\begin{picture}(25,12)

\put(1,0){\vector(1,0){25}}
\put(24,-1){time}

\put(0,8.5){${\cal P}_1$}
\put(0,5.5){${\cal P}_2$}

\put(2.1,-1){\line(0,1){11}}
\put(1,10.3){Mode $\mode_i$ in progress}

\put(2.1,8.3){\fcolorbox{black}{lightgray}{\makebox[3\unitlength][c]{$\tau_{1,1}^{i}$}}}
\put(5.5,8.3){\fcolorbox{black}{lightgray}{\makebox[3\unitlength][c]{$\tau_{4,1}^{i}$}}}
\put(2.1,5.3){\fcolorbox{black}{lightgray}{\makebox[1.5\unitlength][c]{$\tau_{2,1}^{i}$}}}
\put(4,5.3){\fcolorbox{black}{lightgray}{\makebox[3\unitlength][c]{$\tau_{3,1}^{i}$}}}

\put(10.1,8.3){\fcolorbox{black}{lightgray}{\makebox[3\unitlength][c]{$\tau_{1,2}^{i}$}}}
\put(13.5,8.3){\fcolorbox{black}{lightgray}{\makebox[3.1\unitlength][c]{$\tau_{4,2}^{i}$}}}
\put(10.1,5.3){\fcolorbox{black}{lightgray}{\makebox[1.5\unitlength][c]{$\tau_{2,2}^{i}$}}}
\put(12,5.3){\fcolorbox{black}{lightgray}{\makebox[2\unitlength][c]{$\tau_{3,2}^{i}$}}}
\put(11.1,3){\line(0,1){7}}
\put(10.5,10.3){$\operatorname{MCR(j)}$}

\put(17.25,8.3){\fcolorbox{black}{darkgray}{\makebox[5\unitlength][c]{$\tau_{2,1}^{j}$}}}
\put(14.8,5.3){\fcolorbox{black}{darkgray}{\makebox[3\unitlength][c]{$\tau_{1,1}^{j}$}}}
\put(18.55,5.3){\fcolorbox{black}{darkgray}{\makebox[3\unitlength][c]{$\tau_{3,1}^{j}$}}}
\put(17.25,3){\line(0,1){7}}
\put(16.5,10.3){$\mode_j$ begins}

\put(4.5, 2.2){\vector(-1,1){2}}
\put(8, 2.2){\vector(1,1){2}}
\put(10.1,3){\line(0,1){7}}
\put(4,1){{\footnotesize $\substack{\text{release of every}\\ \text{job of $\tau^i$}}$}}

\put(12.5,2.3){\vector(1,1){2}}
\put(10,1){{\footnotesize $\substack{\text{enablement of some} \\ \text{tasks of } \tau^j}$}}

\put(19.6, 2.3){\vector(-1,1){2}}
\put(16.5,1){{\footnotesize $\substack{\text{no more job of } C(i,j) \\ \Rightarrow\ \text{enablement of every task of } \tau^j \\ \text{(end of the transition phase)} }$}}

\end{picture}
}
\caption{Illustration of a mode transition handled by our asynchronous protocol.}
\label{F_futurework}
\end{figure}

Our future work includes:
\begin{enumerate}
\item Establishing a dedicated priority assignment, suitable for scheduling the rem-jobs while minimizing their makespan.
\item Proposing an \emph{asynchronous} protocol \emph{without periodicity} which aims to reduce the enablement delay applied to the new-mode tasks, by enabling them as soon as possible. The main idea of that protocol is depicted in Figure~\ref{F_futurework}. When a processor has no more rem-job to execute, some new-mode tasks are immediately enabled. However, since multiprocessor schedules suffer from scheduling anomalies, immediately enabling the new-mode tasks must be carried out very carefully, in order to not jeopardize the schedulability of the new-mode.
\end{enumerate}

\bibliographystyle{latex8}
\bibliography{energy}

\end{document}